\documentclass[preprint2]{aastex}

\shorttitle{\Herschel\ observations of GX~301--2}
\shortauthors{Servillat et al.}

\def\ebv{$E(B-V)$} 
\def\mum{$\mathrm{\mu}$m} 
\def\msun{$M_{\odot}$} 
\def\lsun{$L_{\odot}$} 
\def\rsun{$R_{\odot}$}

\def\Swift{\emph{Swift}}

\def\Herschel{\emph{Herschel}}
\def\Spitzer{\emph{Spitzer}}
\def\INTEGRAL{\emph{INTEGRAL}}

\begin{document}


\title{\Herschel\ observations of dust around the \\high-mass X-ray binary GX~301--2}


\author{M. Servillat\altaffilmark{1,2,3}, A. Coleiro\altaffilmark{4,2}, S. Chaty\altaffilmark{2,5}, F. Rahoui\altaffilmark{6,3} and J. A. Zurita Heras\altaffilmark{4}}


\altaffiltext{1}{Laboratoire Univers et Th\'eories (CNRS/INSU, Observatoire de Paris, Universit\'e Paris Diderot), 5 place Jules Janssen, 92190 Meudon, France\\ contact email: {mathieu.servillat@obspm.fr}}
\altaffiltext{2}{Laboratoire AIM (CEA/Irfu/SAp, CNRS/INSU, Université Paris Diderot), CEA Saclay, Bat. 709, 91191 Gif-sur-Yvette, France}
\altaffiltext{3}{Harvard University, Department of Astronomy, 60 Garden street, Cambridge, MA, 02138, USA}
\altaffiltext{4}{AstroParticule et Cosmologie (Universit\'e Paris Diderot, CNRS/IN2P3, CEA/DSM, Observatoire de Paris, Sorbonne Paris Cit\'e), 10 rue Alice Domon et L\'eonie Duquet, 75205 Paris Cedex 13, France}
\altaffiltext{5}{Institut Universitaire de France, 103 Boulevard Saint-Michel, 75005 Paris, France}
\altaffiltext{6}{European Southern Observatory, Karl Schwarzschild-Strasse 2, 85748 Garching bei M\"unchen, Germany}


\begin{abstract}
We aim at characterising the structure of the gas and dust around the high mass X-ray binary GX~301--2, a highly obscured X-ray binary hosting a hypergiant star and a neutron star, in order to better constrain its evolution.
We used \Herschel\ PACS to observe GX~301--2 in the far infrared and completed the spectral energy distribution of the source using published data or catalogs, from the optical to the radio range (0.4 to $4\times10^{4}$~\mum). 
GX~301--2 is detected for the first time at 70 and 100~\mum.
We fitted different models of circumstellar environments to the data.
All tested models are statistically acceptable, and consistent with a hypergiant star at $\sim$3~kpc.
We found that the addition of a free-free emission component from the strong stellar wind is required and could dominate the far infrared flux.
Through comparisons with similar systems and discussion on the estimated model parameters, we favour a disk-like circumstellar environment of $\sim$8~AU that would enshroud the binary system. The temperature goes down to $\sim$200~K at the edge of the disk, allowing for dust formation. This disk is probably a rimmed viscous disk with an inner rim at the temperature of the dust sublimation temperature ($\sim$1\,500~K).
The similarities between the hypergiant GX~301--2,  B[e] supergiants and the highly obscured X-ray binaries (in particular IGR~J16318--4848) are strengthened. 
GX~301--2 might represent a transition stage in the evolution of massive stars in binary systems, connecting supergiant B[e] systems to luminous blue variables.
\end{abstract}






\section{Introduction}

Massive stars (more than about eight times the mass of the Sun, e.g. \citealt{Heger:2003p12872}) have been key players in the cosmic history through their enormous input of ionizing photons, energy, momentum, and chemical species into the Universe. 
In addition, the hot, massive stars can episodically form enormous quantities of dust.
The presence of dust and its distribution bring primary information on the evolution of those massive stars \citep[e.g.][]{Kochanek:2011p12617}.

While cool star winds readily form dust (see the review by \citealt{Willson:2000p12849}),
the intense radiation fields of young, massive stars would be expected to prevent dust formation in their circumstellar (CS) environment. 
However, dust is observed around some massive stars, for example in a subset of Carbon Wolf-Rayet stars (WC) containing binaries with colliding winds (e.g. \citealt{Crowther:2007p12079}), in luminous blue variables (LBVs) that show large dusty shells (see the review by \citealt{Humphreys:1994p12826}), or in supergiants showing the B[e] phenomenon (sgB[e], \citealt{Zickgraf:1986p12781,Zickgraf:1998p12657}).
sgB[e] stars are early-type stars with the simultaneous presence of low-excitation forbidden line emission and strong infrared excess in their spectra
\citep{Zickgraf:1998p12657,Lamers:1998p12450}. 
They share some observational properties with LBVs 
which are identified as luminous, hot, unstable supergiants that suffer irregular eruptions like S~Dor and AG~Car or more rarely giant eruptions like P~Cyg and $\eta$~Car \citep{Humphreys:1994p12826}.
All those rare types of stars have in common a rapid mass loss in a dense wind and may represent some of the latest stages of the evolution of massive stars that lead to supernovae \cite[SNe, e.g.][]{Langer:2012p12859}.

It is not always clear whether those different classifications represent stars with different initial conditions, or stars at different evolutionary stages.
It is however clear that dust can be formed around massive stars under specific conditions: strong stellar winds (responsible for high CS densities and thus shielding the dust grains from the stellar ionizing radiation), a high abundance of heavy elements (increasing the probability of dust formation), and clumping of the CS material (locally increasing the density). 
Those conditions are natural around massive ($>20$~\msun) and luminous ($>10^5$ \lsun) objects, such as Wolf-Rayet stars, hot supergiants, and LBVs. 

Another key parameter that could be related to the presence of CS dust around those massive stars may be binarity. 
Indeed, according to \citet{Sana:2012p12865}, binary interaction dominates the evolution of massive stars : over 70\% of all massive stars exchange mass with a companion, leading to a binary merger in one third of the cases.
For sgB[e], it is suggested that the dust can be present in the outer parts of an equatorial disk, that probably forms through the bi-stability mechanism \citep{Lamers:1991p13527} and possibly fast rotation \citep{Bjorkman:1993p13530}.
\cite{Podsiadlowski:2006p13134} noticed that binary mergers produce an initially rapidly rotating merged object that is an excellent candidate for sgB[e] progenitors. 
Furthermore, \cite{Miroshnichenko:2007p13045} proposed that sgB[e] and their low luminosity counterparts (FS~CMa objects) are currently undergoing or have recently undergone a phase of rapid mass exchange in a binary system, associated with a strong mass loss and dust formation. 
Mass transfer in close binary systems with massive stars could then have an important role in shaping the structure of the diffuse CS environment \citep[e.g.][]{Plets:1995p12556,Millour:2011p13762}.


In many cases, the inferred companion star could be much fainter than the massive primary and may have remained undetected. Among binary stars with a massive primary, high-mass X-ray binaries (HMXBs) are interesting test cases as the (pulsed) X-ray emission provides detailed information on the binary composition, orbital parameters and environment. 
Those may represent a new stage in the complex evolution of massive stars.
In particular, a growing number of highly obscured HMXBs has been revealed by the \INTEGRAL\ hard X-ray observatory (\citealt{Matt:2003p13015,Filliatre:2004p11824,Walter:2006p9106,Chaty:2012p9588,Coleiro:2013p10853}, see also \citealt{Chaty:2013p13738} for a recent review). Those systems may harbour super or hypergiant stars in binaries with a compact object and enshrouded in a dense CS environment, probably hosting dust.
We performed \Herschel\ observations in the mid-infrared of six supergiant HMXBs in order to test the presence of dust and better understand its structure in those objects \citep[see preliminary results in][]{Chaty:2013p11594}, and focus here on GX~301--2.


GX~301--2 is an obscured HMXB system
consisting of an accreting neutron star fed by the stellar wind of a blue hypergiant (HG) B1\,Ia$^+$ star (catalogued as Wray 977, or BP~Cru, see e.g. \citealt{Lewin:1971p13042,Jones:1974p12942,Kaper:1995p13145,Kaper:2006p11663}). 
The absorption varies from $10^{22}$ to $10^{24}$~atoms~cm$^{-2}$ \citep{Mukherjee:2004p12588} while the reddening is around \ebv =1.96 \citep{Kaper:2006p11663}.
The system has an orbital period of $\sim$41.5 days with an eccentricity of 0.46 \citep{Koh:1997p12328}, and
GX~301--2 is one of the slowest known pulsars with a period that slowly varied between 675 and 700~s \citep{White:1976p12648,Pravdo:1995p12853,Evangelista:2010p13139}.
The magnetic field is around $4\times10^{12}$~G, indicating a classical pulsar \citep{Kreykenbohm:2004p13538}.
The X-ray mass function of the system has been estimated to be $\sim$31.9~\msun\ \citep{Sato:1986p13226}, the highest known for a HMXB with a pulsar companion, giving a mass of $43\pm10$ ~\msun\ for the companion star and $1.85\pm0.6$~\msun\ for the neutron star \citep{Kaper:2006p11663}.
The X-ray light curve is characterized by bright flares occurring just before periastron passage of the neutron star \citep{Watson:1982p12647}. 
Those are clearly visible in the X-ray light-curve\footnote{generated with the online tool on http://www.swift.ac.uk/ \citep{Evans:2007p6460}} from the \Swift\ Burst Alert Telescope (see Figure~\ref{xlc}). 
The distance of GX~301--2 has been estimated to be $3.1\pm0.64$~kpc using a spectral energy distribution fitting procedure \citep{Coleiro:2013p10853}, consistent with the 3--4~kpc range proposed earlier \citep{Kaper:2006p11663}.
{Therefore the line of sight crosses the Southern Coalsack (e.g. Kaper et al. 2006) and passes through one or more spiral arms in our Galaxy, potentially contaminating the emission.}

\cite{Kaplan:2006p11690} and \cite{Moon:2007p11833} studied the CS environment of the system and reported the presence of dust around the system through the detection of silicate absorption features and continuum components with low temperatures. However, they do not explore the geometry of this CS component.
\cite{Moon:2007p11833} noticed the presence of low ionization potential forbidden emission lines, and discuss similarities with LBVs in the mid-infrared spectra.



In this paper we report on new \Herschel\ observations of GX~301--2 (Section~2) and model the emission of the source in the wavelength domain from 0.4 to $4\times10^{4}$~\mum\ in order to characterize the geometry of the CS environment. We detail the models used in Section~3, give the results of the fits to the data in Section~4 and discuss the properties and the possible geometries of the CS environment of GX~301--2 in Section~5.


\section{Data acquisition and processing}



   \begin{figure}[t]
   \centering
   \includegraphics[width=\hsize]{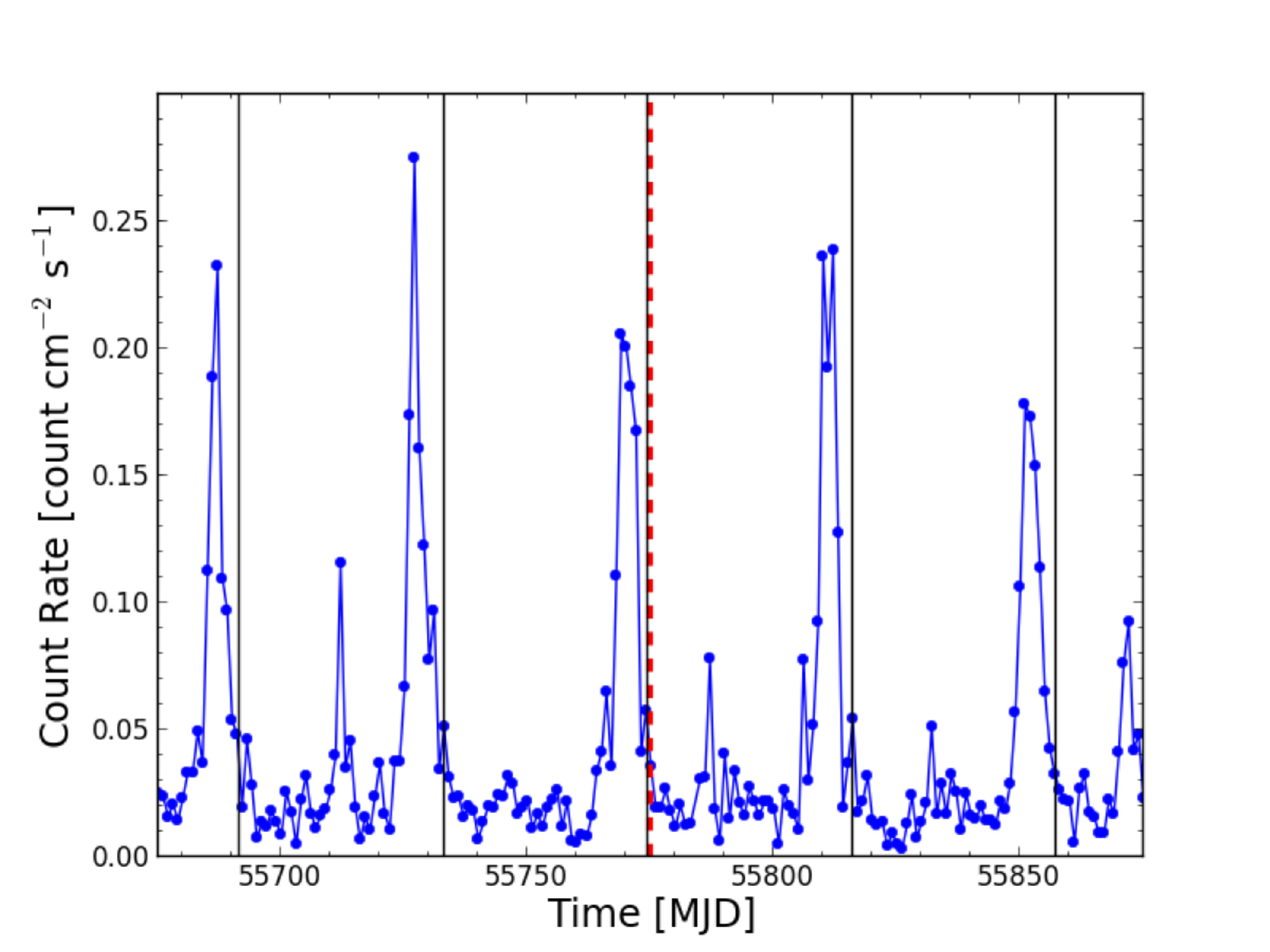}
      \caption{X-ray lightcurve of GX~301--2 with \Swift/BAT (15--150~keV). The vertical lines indicate the periastron passages (based on \citealt{Koh:1997p12328}) and the red dashed vertical line the epoch of the \Herschel/PACS observations on MJD 55775.05, right after periastron.}
         \label{xlc}
   \end{figure}

   \begin{figure}[t]
   \centering
   \includegraphics[width=\hsize]{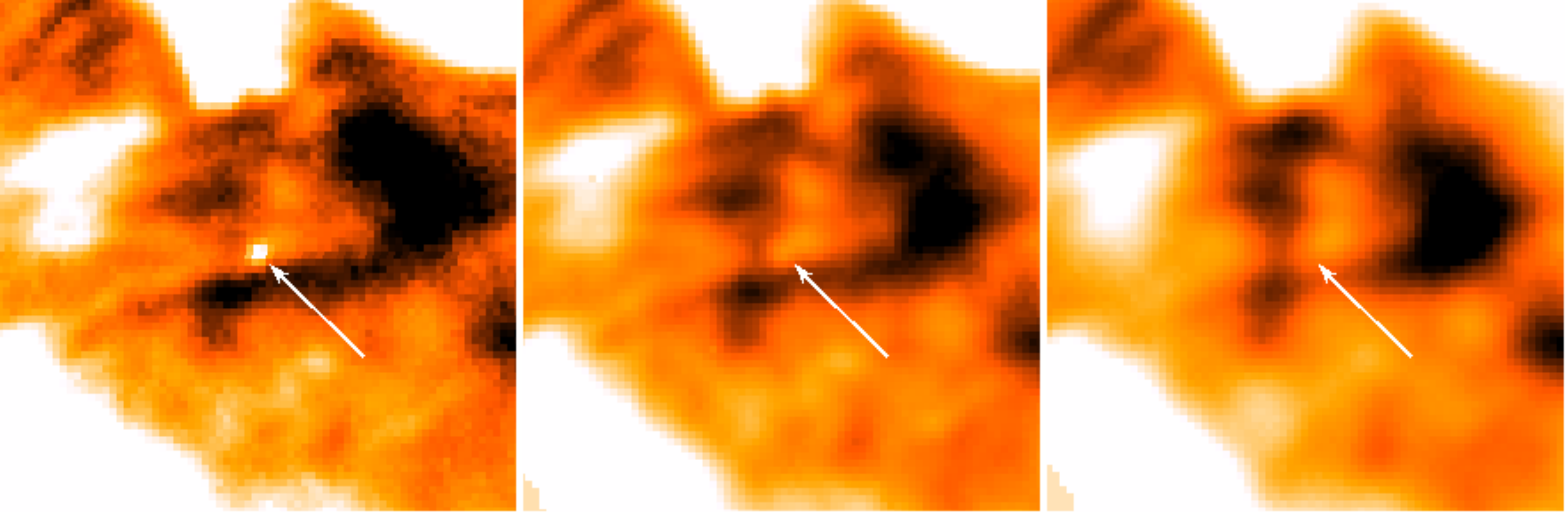}
      \caption{\Herschel/PACS images centered on GX~301--2 (indicated by an arrow). From left to right, the blue, green and red bands are shown (70, 100 and 160~\mum). The size of the snapshots is $2.5\arcmin\times2.5\arcmin$,{ North is up, East is left.}}
         \label{img_herschel}
   \end{figure}

\begin{deluxetable}{cccccc}
\tabletypesize{\normalsize}
\tablecaption{Fluxes in optical to radio bands\label{fit_data}}
\tablewidth{0pt}
\tablehead{
\colhead{Band name}          & \colhead{Wavelength}     & \colhead{Band width}        & \colhead{Flux}          & \colhead{Flux error}         & \colhead{Note} \\
               & ($\mu$m)        & ($\mu$m)              & (Jy)          & (Jy)      & 
}
\startdata 
 B & 0.441 & 0.102 & 0.034 & 0.012 & Tycho-2, \citet{Hg:2000p12294} \\ 
 V &  0.551 & 0.089 & 0.206 & 0.014 & Tycho-2, \citet{Hg:2000p12294} \\
 J &  1.242 & 0.240 & 3.28 & 0.07 & 2MASS, \citet{Skrutskie:2006p4390} \\
 H &  1.651 & 0.267 & 3.80 & 0.10 & 2MASS, \citet{Skrutskie:2006p4390} \\
 Ks &  2.166 & 0.282 & 3.59 & 0.07 & 2MASS, \citet{Skrutskie:2006p4390} \\
 WISE-1$^*$ &  3.353 & 0.663 & 0.47 & \nodata & Saturated, \citet{Wright:2010p12301} \\
 IRAC-1 & 3.6 & 0.77 & 2.0 & 0.2 & \citet{Kaplan:2006p11690} \\
 IRAC-2 & 4.5 & 1.06 & 1.6 & 0.2 & \citet{Kaplan:2006p11690} \\
 WISE-2$^*$ &  4.603 & 1.042 & 0.70 & \nodata & Saturated, \citet{Wright:2010p12301} \\
 IRAC-3 & 5.8 & 1.43 & 1.16 & 0.12 & \citet{Kaplan:2006p11690} \\
 ISO-LW2$^*$ & 6.75 & \nodata & 0.837 & \nodata & \citet{Kaper:2006p11663} \\
 IRAC-4 & 8.0 & 2.9 & 0.78 & 0.08 & \citet{Kaplan:2006p11690} \\
 MSX-A &  8.28 & 3.36 & 0.63 & 0.07 & \citet{Price:2001p12314} \\
 AK09$^*$ & 9.0 & 4.9 & 0.60 & 0.02 & \citet{Ishihara:2010p14025} \\
 ISO-LW10$^*$ & 11.5 & \nodata & 0.517 & \nodata & \citet{Kaper:2006p11663} \\
 WISE-3 & 11.560 & 5.507 & 0.41 & 0.01 & \citet{Wright:2010p12301} \\
 MSX-C & 12.13 & 1.72 & 0.45 & 0.05 & \citet{Price:2001p12314} \\
 MSX-D & 14.65 & 2.23 & 0.32 & 0.03 & \citet{Price:2001p12314} \\
 AK18$^*$ & 18.0 & 11.7 & 0.20 & 0.03 & \citet{Ishihara:2010p14025}\\
 WISE-4 & 22.088 & 4.101 & 0.22 & 0.01 & \citet{Wright:2010p12301} \\
 ISO-PHT3$^*$ & 25 & \nodata & 0.089 & \nodata & \citet{Kaper:2006p11663} \\
 PACS-B & 70 & 20 &  0.061 & 0.0066 & This work \\
 PACS-G & 100 & 40 & 0.061 & 0.0149 & This work \\
 8.6 GHz & 34860 & 600 & 0.72$\times10^{-3}$ & 0.16$\times10^{-3}$ & \citet{Pestalozzi:2009p12072} \\
\enddata
\tablecomments{See the references in the text.\\$^*$ Those points were not used in the modelling as they have lower precision (due to saturation, large bandwidth or large errors).}
\end{deluxetable}

We performed sensitive far-infrared (60--210~\mum) observations\footnote{The \mbox{OBSIDs} are 1342225124, 1342225125, 1342225126 and 1342225127} of GX~301--2 on 2011 August 2 with the ESA \Herschel\ Space Observatory \citep{Pilbratt:2010p13619}, in particular employing \Herschel 's large telescope and powerful science payload to do photometry using the PACS instrument \citep{Poglitsch:2010p13601}. 
At the date of the observations, the binary system was at its periastron, right after the periodic X-ray flare that can be seen in Figure~\ref{xlc}.
Observations were performed in the 3 available bands (blue 60-85 $\mu$m, green 85-130 $\mu$m and red 130-210 $\mu$m) in mini-scan map mode: medium speed, 10 scan legs of 2.5\arcmin\ length with 2.0\arcsec\ cross-scan step, with orientation angles at 70 and 110 degrees, and a repetition factor of 5 and 10 for each orientation angle for the blue/red and green/red filters respectively. 
This leads to a total integration time of 800, 1600 and 2400~s in the blue, green and red band, respectively.

{The images are shown in Figure~\ref{img_herschel} where a point source corresponding to GX~301--2 can be seen in the center of the frame. The point source appears to be surrounded by an extended structure. Based on IRAS observations, \citet{Huthoff:2002p14022} suggested that the extended structure could be a wind bow-shock (see their Fig. 4). We further analyse the extended emission and possible cavity around GX 301--2 in a companion paper (Coleiro et al. in preparation).}

We used HIPE \citep{Ott:2010p10804} to reprocess the data with custom scripts for PACS (M. Sauvage, private communication). 
The software \textit{getsources} \citep{Menshchikov:2012p9675} was then used to detect the sources and filaments, and to extract fluxes using a multi-wavelength and multi-scale method.
As the source is faint and the structure of the background complex, we masked the area outside a $2.4\arcmin\times1.7\arcmin$ region around the target to avoid the presence of extended features seen at the edges of the images in Figure~\ref{img_herschel}.
We detected a point source at the position of GX~301--2 in the blue and green bands with fluxes of $61.0\pm6.6$ and $61.2\pm14.9$~mJy (with significances of 15.4 and 3.8 sigma respectively). No significant detection was made for the red band with a detection limit of 250~mJy at 3 sigma (note the presence of a close-by extended source blended with the point source).

The optical-infrared spectral energy distribution (SED) was completed using $B$ and $V$-band magnitudes from the Tycho-2 catalog \citep{Hg:2000p12294}, and $JHKs$ magnitudes from the Two Micron All Sky Survey (2MASS, \citealt{Skrutskie:2006p4390}). We included fluxes at 3.6, 4.5, 5.8 and 8.0~\mum\ from the Galactic Legacy Infrared Midplane Survey Extraordinaire (GLIMPSE) catalog based on observations with IRAC onboard \Spitzer\ \citep{Benjamin:2003p12299,Churchwell:2009p12300}. As the central pixels of the source were saturated at 3.6 and 4.5~\mum, we used the flux values estimated by \citet{Kaplan:2006p11690}. We finally included fluxes at 12 and 22~\mum\ obtained with the \emph{Wide-field Infrared Survey Explorer} (WISE, \citealt{Wright:2010p12301}) and fluxes at 8.28, 12.13 et 14.65~\mum\ obtained with the \emph{Midcourse Space Experiment} (MSX, \citealt{Price:2001p12314}). 
{The flux measurements from various observatories are reported in Table~\ref{fit_data}. Within the errors, the infrared measurements with different instruments are consistent. We thus used all these data simultaneously to characterize the emission of the source.}

GX~301--2 was also observed at radio wavelengths \citep{Pestalozzi:2009p12072} with ATCA (4.8 and 8.6~GHz) at 12 epochs spread between November 2008 and February 2009. 
\citet{Pestalozzi:2009p12072} found that the radio source showed a negative index during the X-ray outburst, and a positive index otherwise, though with large error bars.
They suggest that the radio emission originates in two components: a persistent thermal emission from the wind of the HG mass donor Wray 977, and an episodic emission, perhaps a weak jet, that appears at the time of the X-ray outburst. 
The mean flux at 8.6~GHz when the source is detected is $0.72\pm0.16$~mJy.
{To estimate the flux due to the presence of strong winds around massive stars, we used Equation~4 in \cite{Scuderi:1998p12075} and input parameters estimated by \citet{Kaper:2006p11663} (i.e. a mass loss rate of $\dot{M}=10^{-5}$~\msun~yr$^{-1}$,  temperature $T = 18\,100$~K, and terminal velocity of the wind $v_{\inf} = 305$~km~s$^{-1}$). We thus expect a 8.6~GHz flux of 2.5~mJy for \mbox{GX~301--2}. The discrepancy possibly indicates a lower mass-loss rate or larger distance, but the radio flux from a stellar wind can naturally vary by a factor of a few (Scuderi et al. 1998).}
During our \Herschel\ observations, we can nevertheless expect free-free emission from the stellar wind, which may have some contribution to the mid-infrared emission down to $\sim$10~\mum\ with a positive spectral index $\alpha=0.6$ \citep[$S_{\nu}\propto\nu^{\alpha}$,][]{Wright:1975p12076}.


\section{Modeling the emission}
\label{models}



The collected data allowed us to fit spectral models over five orders of magnitude, from 0.4 to $4\times 10^{4}$~\mum, with \Herschel\ data in the 60--210~\mum\ range bridging the gap between mid-infrared and radio observations for the first time for this kind of object. 

We used the stellar photosphere model of \cite{Castelli:2004p12382}, hereafter CK04, and the stellar parameters derived by \cite{Kaper:2006p11663} for the star Wray 977 associated to GX~301--2 are: $T = 18\,100$~K, $\mathrm{log}(g)=2.38$ and $R = 70$~\rsun, for a distance $D=3040$~pc. We thus selected the closest available CK04 model with $T = 18\,000$~K and $\mathrm{log}(g)=2.5$ and kept the radius of the star fixed.

We also built an extinction law following \cite{Cardelli:1989p11745} in the visible, \cite{Indebetouw:2005p11602} between 1.25 and 8.0~\mum, \cite{Lutz:1996p11644} between 8.0 and 24~\mum\ and \cite{Moneti:2001p11726} above 24~\mum. 
The main parameter for the extinction law is the excess color \ebv. We adopt \mbox{$R_{V}= A_{V} / $\ebv\ $ = 3.1$} (see e.g. \citealt{Savage:1979p5573}). 
As the reddening is high for GX~301--2, the extinction law has a large impact on the shape of the SED. 
In particular, broad silicate absorption lines around 10 and 20 \mum\ are visible and taken into account in the extinction law \citep{Lutz:1996p11644}.
The obtained extinction law is particularly well suited for the case of GX~301--2, being largely based on observations in the Galactic Center and the Galactic plane. However, there could be local fluctuations of this law that are still not properly measured.


The optical-infrared SED of GX~301--2 has already been fitted between 1 and 10~\mum\ by \citet{Kaplan:2006p11690}, using two components: the star and a spherical black body emission with temperature 740~K and an extension $R_{dust} = 9 \times R_{star}$ to model the surrounding dust. Using Spitzer data up to 38~\mum, \citet{Moon:2007p11833} found that multiple emission components were required to model the continuum emission of the dust (a hot component with $T\sim700$~K and a warm component with $T\sim180$~K).

In order to constrain the geometry of the CS environment, we considered several models that could explain the infrared excess found in this source: (i) a power law spanning several orders of magnitude, (ii) spherical black body dust components (as in \citealt{Kaplan:2006p11690} and \citealt{Moon:2007p11833}), (iii) disk-like dust components. For (ii) and (iii), we also tested the addition of a power law with a fixed spectral index $\alpha=0.6$ and the observed flux at 8.6~GHz in order to model the free-free emission from the stellar wind  detected by \citet{Pestalozzi:2009p12072}.

GX~301--2 appears to be similar to two obscured X-ray binary systems hosting a sgB[e] star: IGR~J16318-4848 \citep{Revnivtsev:2003p12374,Filliatre:2004p11824,Kaplan:2006p11690,Moon:2007p11833,Chaty:2012p9588} that present a circumstellar disk with dust, and XTE J0421+560 (CI Cam) for which IRAS 12--100~\mum\ data suggested the existence of a substantial circumstellar dust shell \citep{Belloni:1999p13664,Clark:1999p13701}.
We thus investigate the possibility of having a similar distribution of dust as a disk surrounding the hyper giant star, as it is generally observed for sgB[e].
\cite{Lamers:1998p12450} pointed out that the CS envelope geometry is most probably disk-like for those stars, which naturally provides conditions for a high density of the CS material and, as a consequence, for shielding of CS dust from the ionizing stellar radiation.
We thus used a similar model to the one proposed by \citet{Chaty:2012p9588} for IGR~J16318-4848, initially based on the dust structure of Herbig Ae/Be stars \citep{Monnier:2005p12460}.
The CS environment of those objects is well described as a simple disk model possessing a central optically thin (dust-free) cavity, ringed by hot dust --- a rim --- emitting at the expected sublimation temperature ($\sim$1\,500~K), hereafter called a \emph{rimmed disk}.
For some systems, however, the inner gas in the mid-plane may be optically thick, partially shielding the innermost dust from stellar radiation and causing the dust sublimation radius to shrink for the same sublimation temperature. This would correspond to a \emph{classical disk} model.

We assumed that the disk is completely flat with a ring-like rim of constant temperature $T_{\mathrm{rim}}$, located at an inner radius $R_{\mathrm{in}}$ and of width $H_{\mathrm{rim}}$, as in  \cite{Lachaume:2007p11825}.
The equation used to model the flux can therefore be written as follows:
\begin{eqnarray}
F_\nu(\nu) &=& \left(\frac{R_*}{D}\right)^2\times \mathrm{CK04}(\nu,T_*) \nonumber\\
           &+& {2\pi\frac{H_{\mathrm{rim}}R_{\mathrm{in}}}{D^2}\mathrm{cos}(i)B(\nu,T_{\mathrm{rim}})} \nonumber\\ 
           &+& { 2\pi\frac{\mathrm{cos}(i)}{D^2}\int^{R_{\mathrm{out}}}_{R_{\mathrm{in}}}rB(\nu,T(r))\mathrm{d}r}
\label{model_rimmeddisk}
\end{eqnarray}

where $T_*$, $R_*$ et $D$ are respectively the temperature, the radius and the distance of the companion star, and CK04 is the stellar photosphere model (\citealt{Castelli:2004p12382}).
$B(\nu,T)$ is the Planck function at the frequency $\nu$.
The disk is defined by $i$, $T_{\mathrm{in}}$, $R_{\mathrm{out}}$ which are respectively the inclination of the disk, the temperature at the inner radius $R_{\mathrm{in}}$, and the outer radius.
$T(r)=T_{\mathrm{in}}\left(r/R_{\mathrm{in}}\right)^{-q}$ is the disk temperature at a given radius $r$ where q is a dimensionless parameter generally ranging from 0.5 to 0.75 (irradiated to viscous disc, \citealt{Chiang:1997p12390}).

We fixed the inclination to 60$^{\circ}$ as estimated by \cite{Kaper:2006p11663} for the inclination of the binary system, assuming that the inclination of a disk around the companion star would be similar. We only considered the case of a viscous disk ($q=0.75$), as expected for this kind of system (e.g. \citealt{Okazaki:2007p13772}).
It is clear from Equation~\ref{model_rimmeddisk} that we expect degeneracies in the fits between the radii $R_{*}$, $R_{\mathrm{in}}$ and the distance $D$, so the absolute value of those parameters should be taken with care. There is also a degeneracy between $H_{\mathrm{rim}}$ and $R_{\mathrm{in}}$, we thus arbitrarily fixed $H_{\mathrm{rim}}$ to 15~\rsun\ following the results of \cite{Chaty:2012p9588}.


%

%


\section{Fit results}

   \begin{figure}[b!]
   \centering
   \includegraphics[width=\columnwidth]{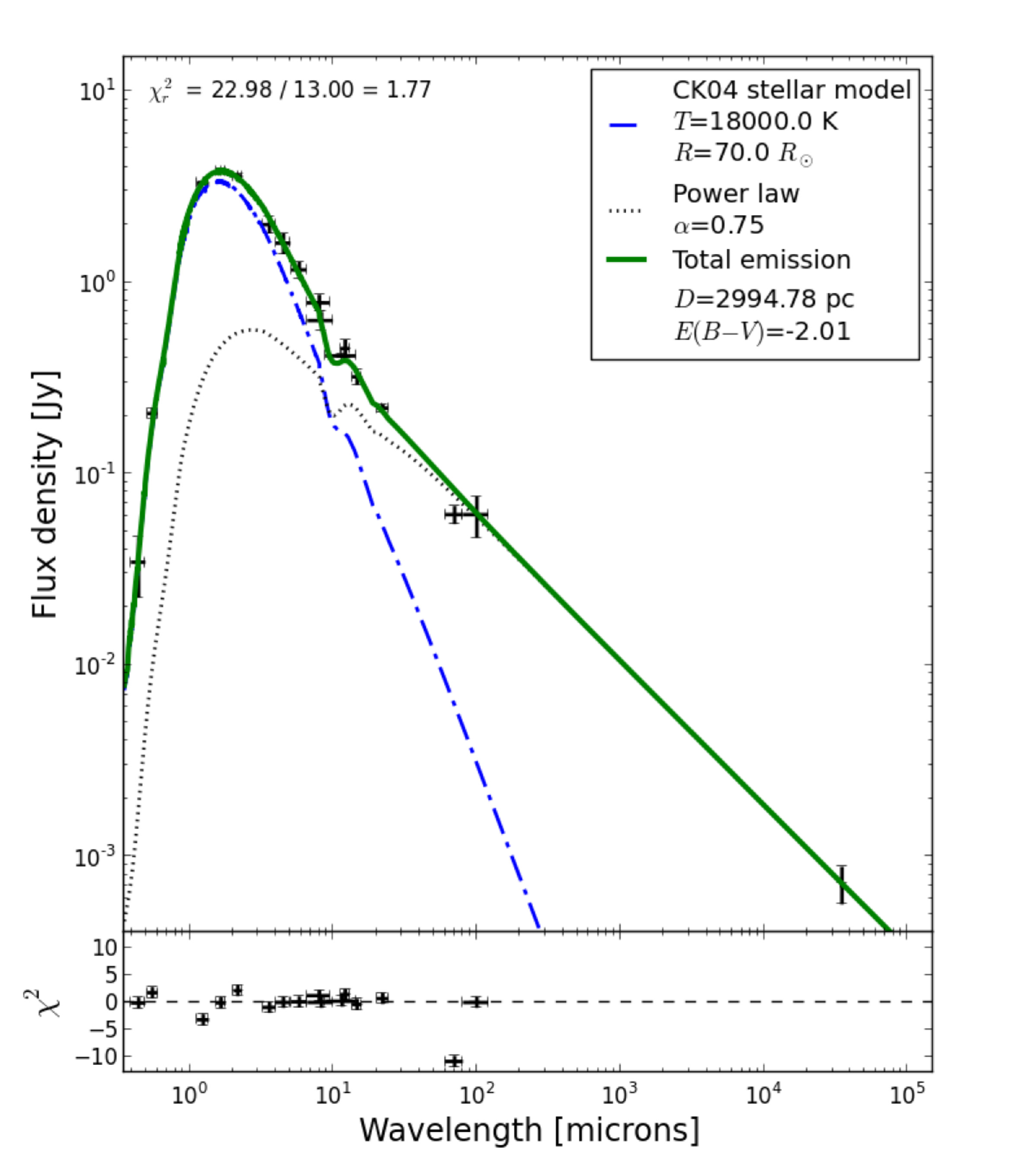}
   \caption{SED of GX~301--2 (Wray 977) fitted with a stellar photosphere model and a power law to account for the infrared excess.}
   \label{sed0}%
   \end{figure}


\begin{deluxetable}{ccccccccc}
\tabletypesize{\scriptsize}
\tablecaption{Fit results\label{fit_results}}
\tablewidth{0pt}
\tablehead{
\colhead{Model}          & \colhead{$D$}     & \colhead{\ebv}        & \colhead{$T_\mathrm{in}$}          & \colhead{$R_\mathrm{in}$}         & \colhead{$R_\mathrm{out}$}          & \colhead{$T_\mathrm{rim}$}          & \colhead{$\alpha$} & \colhead{$\chi^2$/dof} \\
               & (pc)        &               & (K)          & (\rsun)      & (\rsun)       & (K)           &       & 
}
\startdata 
Power law & 2994$\pm$54  & 2.01$\pm$0.04 & -- & --        & -- & --            & 0.75$\pm$0.01   & 23.0/13 \\
[0.5ex] \tableline \\[-2ex]
Sphere         & 2888$\pm$62  & 2.08$\pm$0.06 &  411$\pm$47  & --           & 1288$\pm$197  & --            & --    & 47.4/12 \\
               & 2950$\pm$48  & 2.06$\pm$0.04 &  470$\pm$69  & --           &  919$\pm$169  & --            & [0.6] & 25.9/12\\
[0.5ex] \tableline \\[-2ex]
Classical disk & 2784$\pm$42  & 2.05$\pm$0.04 & 1658$\pm$127 & [70]         & 4263$\pm$1437 & --            & --    & 21.4/12 \\
               & 2830$\pm$42  & 2.04$\pm$0.04 & 1711$\pm$150 & [70]         & 1785$\pm$623  & --            & [0.6] & 18.6/12 \\
[0.5ex] \tableline \\[-2ex]
Rimmed disk    & 2961$\pm$95  & 1.97$\pm$0.05 &  238$\pm$35  & 3275$\pm$817 & 4208$\pm$802  & 1837$\pm$306  & --    & 14.1/10 \\
               & 3215$\pm$405 & 1.92$\pm$0.09 &  400$\pm$95  & 1129$\pm$419 & 1858$\pm$339  & 3267$\pm$1672 & [0.6] &  ~~8.4/10 \\
\enddata
\tablecomments{The parameters reported here are defined is section~\ref{models}. The star temperature $T_{*}=18\,000$~K, and radius $R_{*}=70$~\rsun\ are fixed parameters. When other parameters are fixed, we indicate the value in brackets. $R_\mathrm{rim}$~is equal to $R_\mathrm{in}$ and $H_\mathrm{rim}$ is fixed to 15~\rsun\ following \cite{Chaty:2012p9588}. The inclination is fixed to $i=60^{\circ}$ \citep{Kaper:2006p11663}.}
\end{deluxetable}


   \begin{figure*}
   \centering
   \includegraphics[width=\columnwidth]{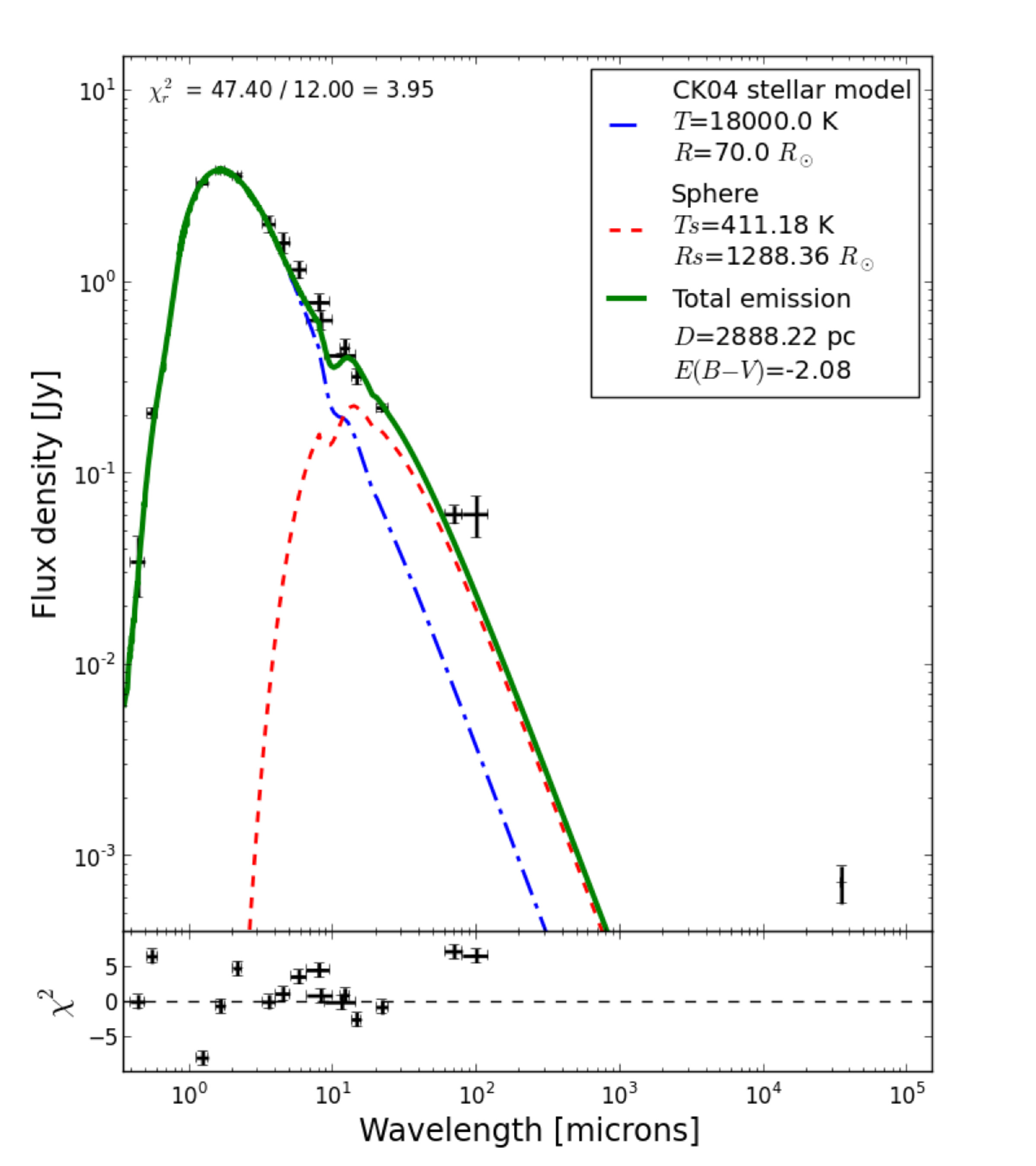}
   \includegraphics[width=\columnwidth]{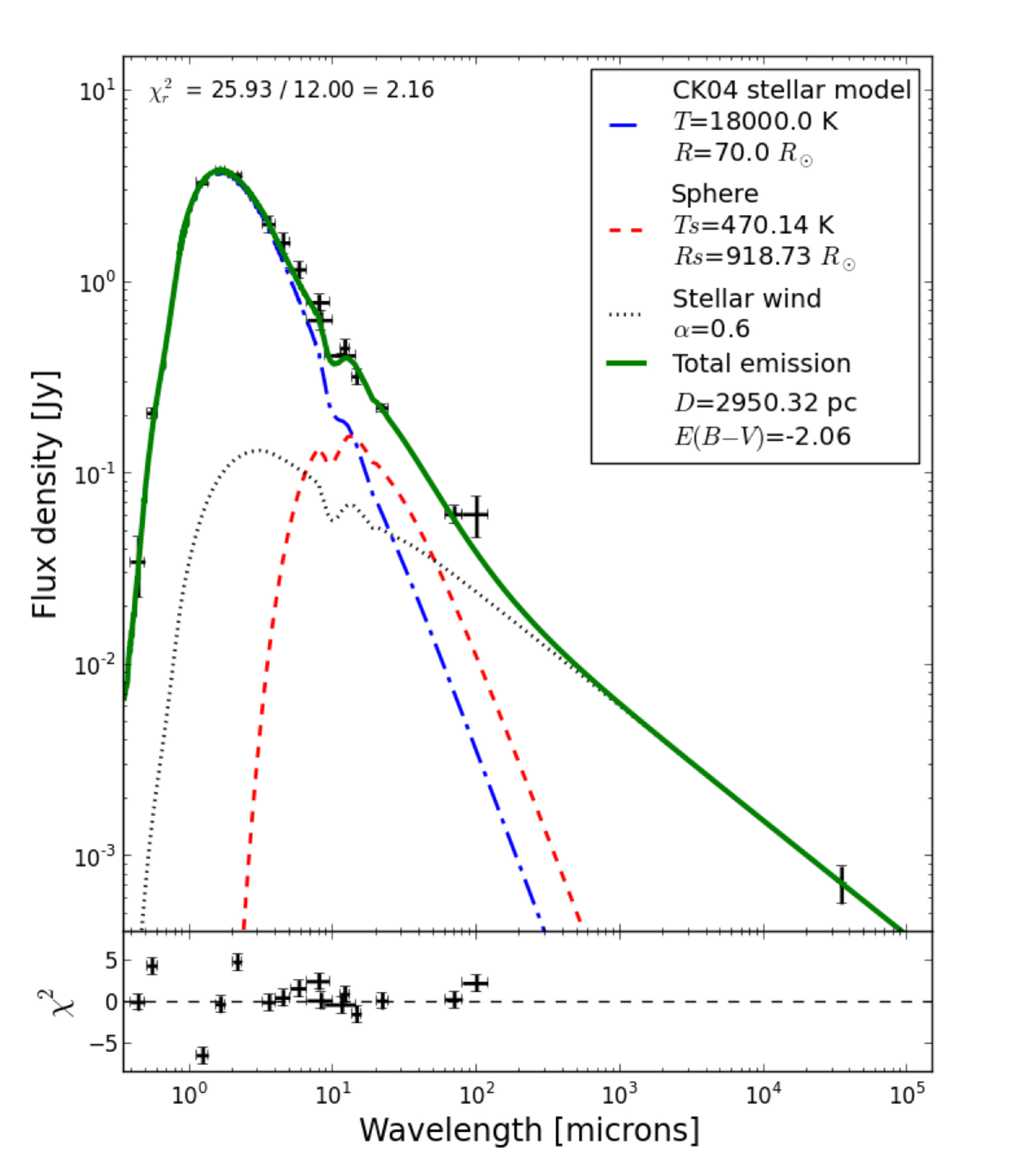}
   \caption{SED of Wray 977 (GX~301--2) fitted with a stellar photosphere model and a spherical black-body component (left) and an additive power-law to account for the stellar wind (right).}
   \label{sed1}%
   \end{figure*}

   \begin{figure*}
   \centering
   \includegraphics[width=\columnwidth]{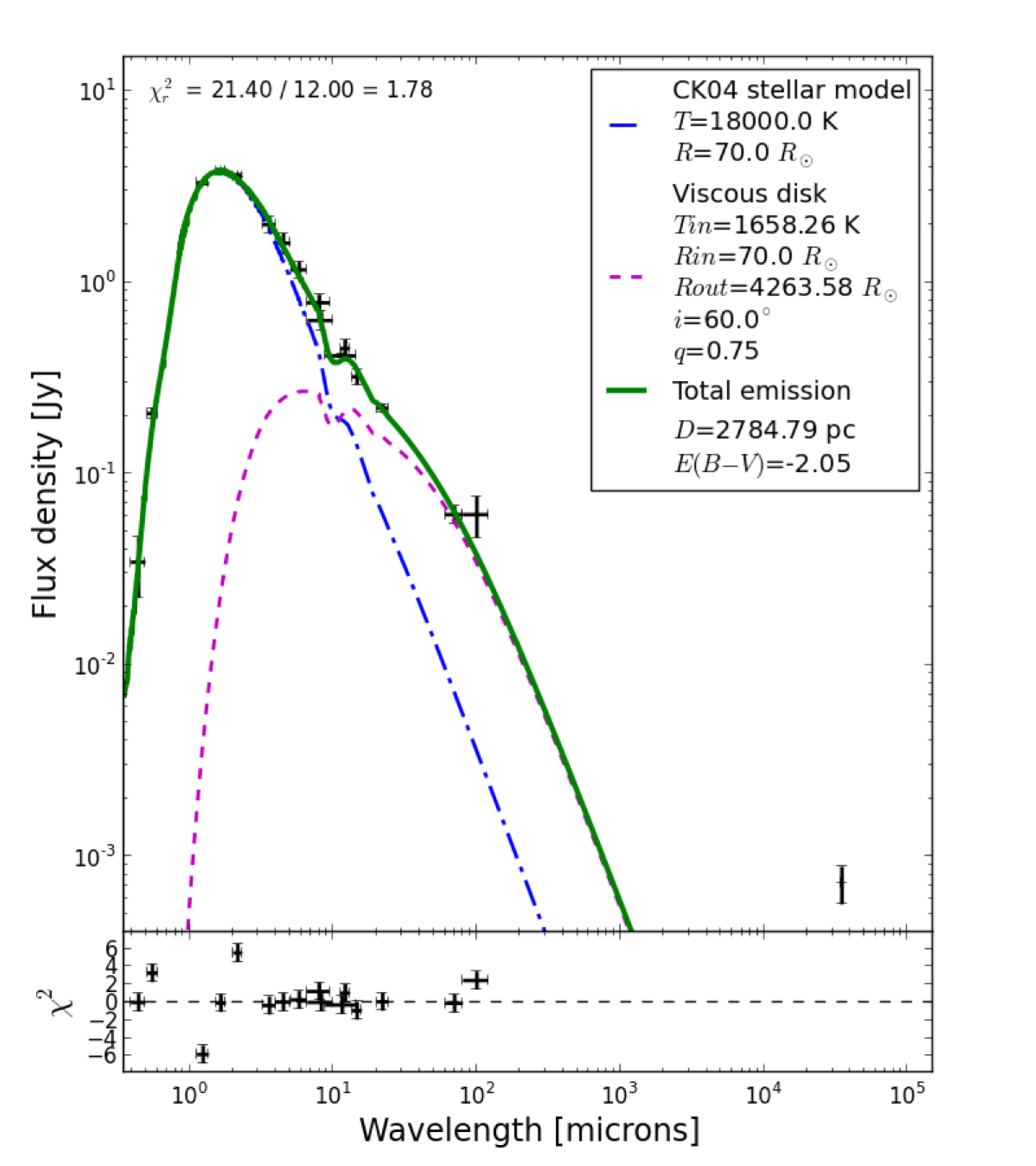}
   \includegraphics[width=\columnwidth]{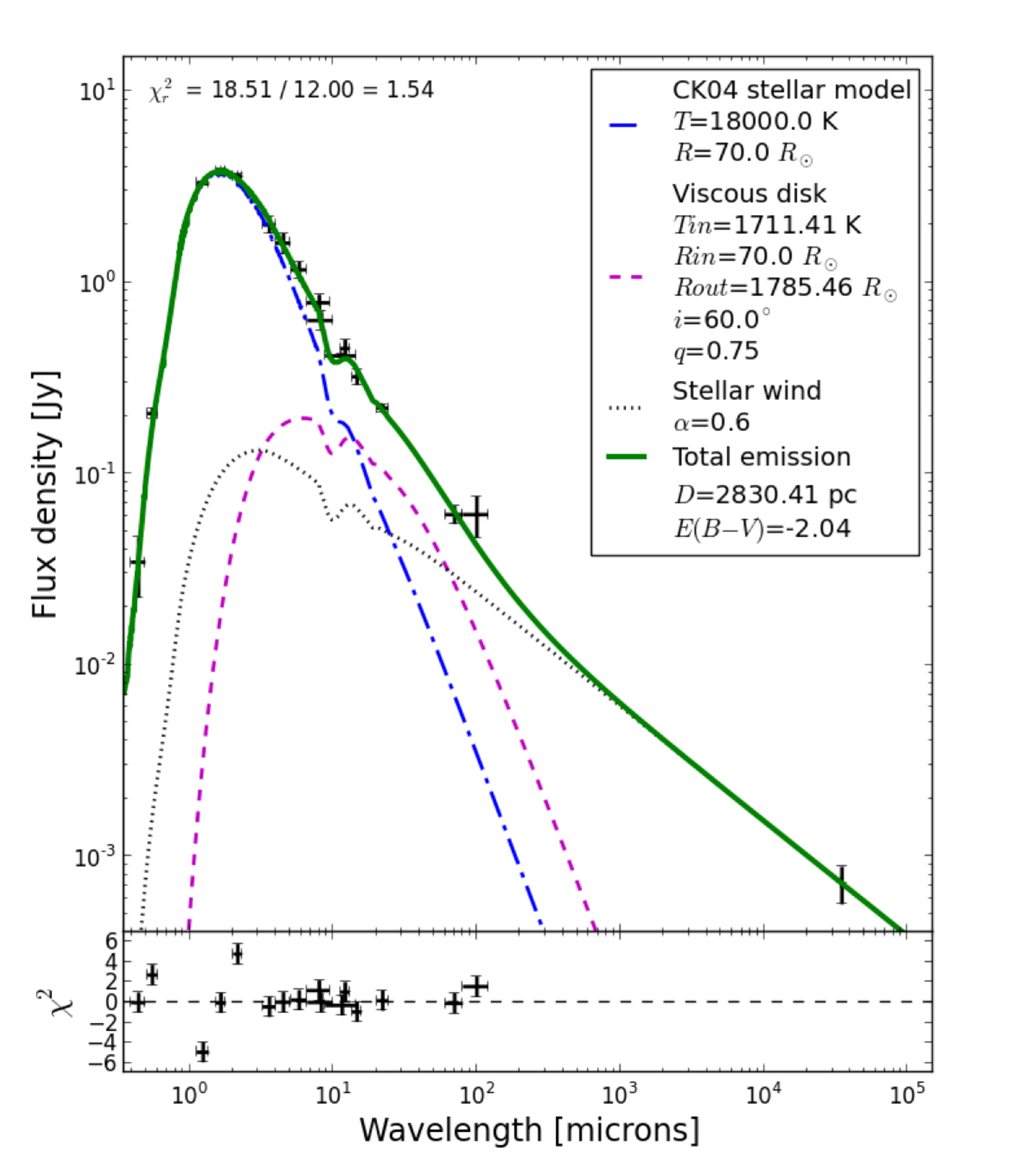}
   \caption{SED of Wray 977 (GX~301--2) fitted with a stellar photosphere model and a classical disk component (left) and an additive power-law to account for the stellar wind (right).}
   \label{sed2}%
   \end{figure*}

   \begin{figure*}
   \centering
   \includegraphics[width=\columnwidth]{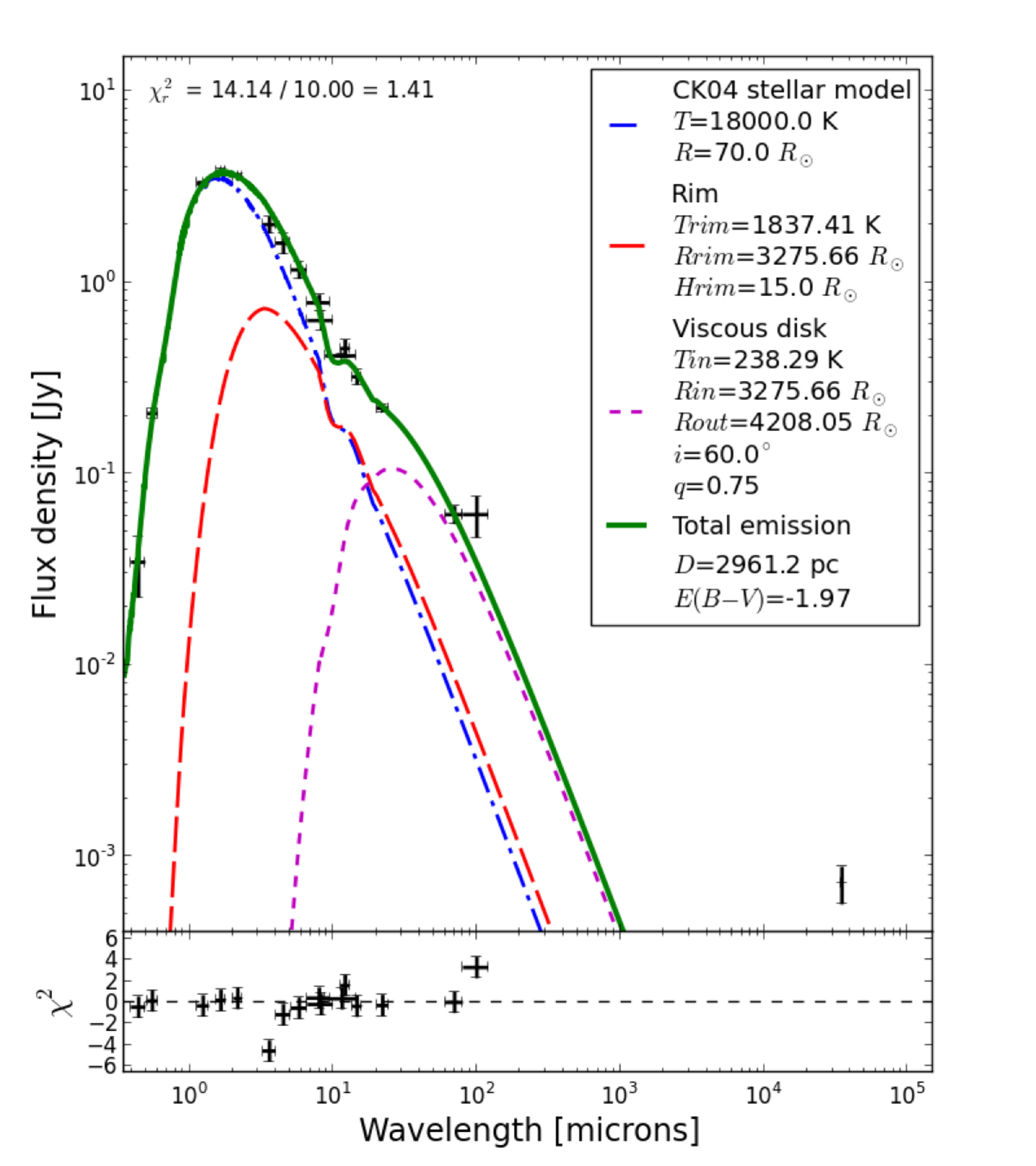}
   \includegraphics[width=\columnwidth]{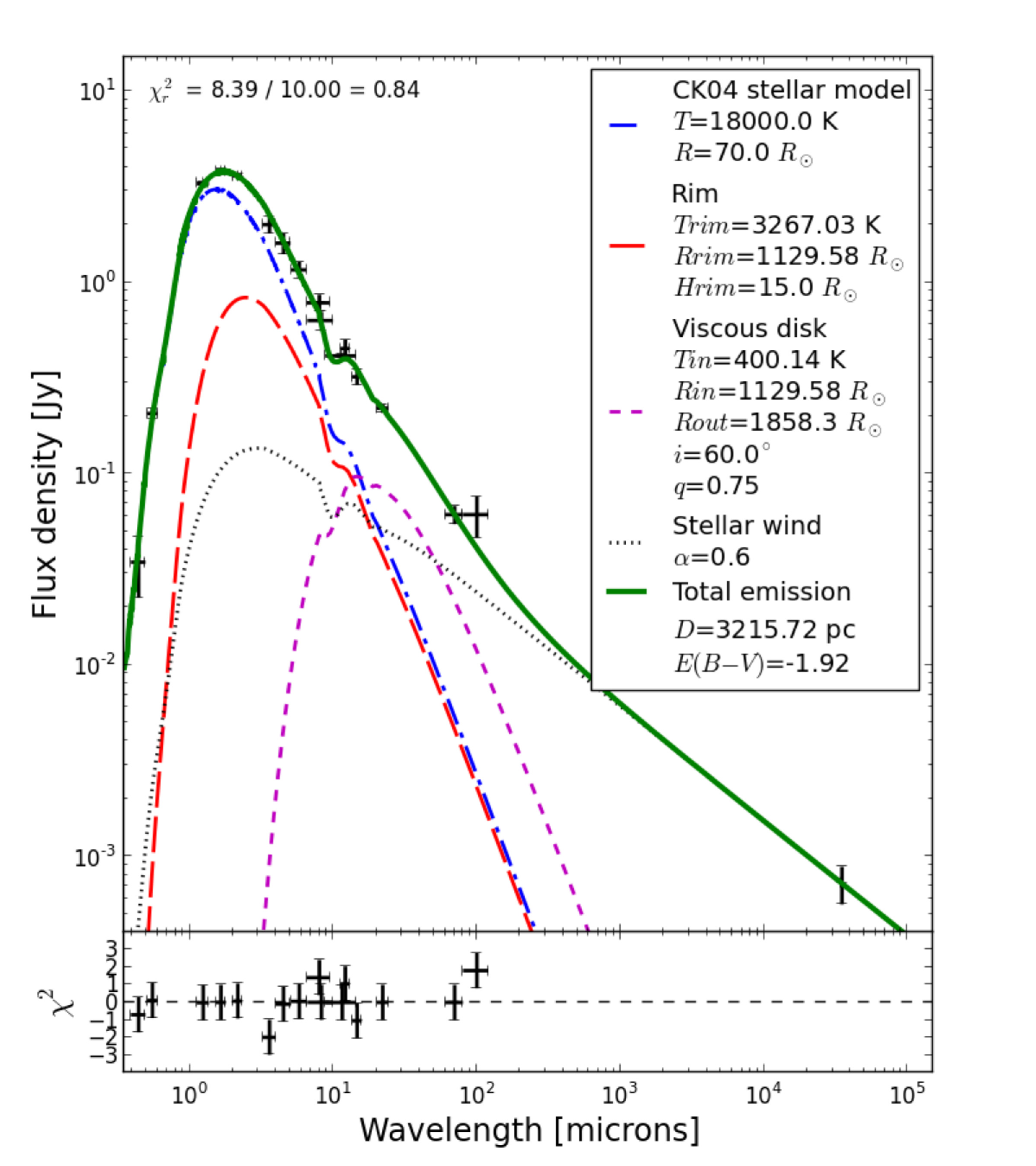}
   \caption{SED of Wray 977 (GX~301--2) fitted with a stellar photosphere model and a rimmed disk component (left) and an additive power-law to account for the stellar wind (right).}
   \label{sed3}%
   \end{figure*}


We report the results of the SED fitting in Table~\ref{fit_results} and plot the data and models in Figures~\ref{sed0} to~\ref{sed3}.

We first tried to fit a power law in addition to the CK04 stellar photosphere model (and including the extinction law). We found a statistically acceptable fit with an index $\alpha=0.75\pm0.01$ (see Table~\ref{fit_results}). This index is different from the 0.6 expected for stellar winds \citep{Wright:1975p12076}. There is a deviation from the model in one \Herschel\ energy band (accounting for most of the $\chi^{2}$ excess, see Figure~\ref{sed0}) that may indicate a marked curvature in the spectrum, and the necessity to explore more complex models.

In order to account for the presence of dust around the system (see \citealt{Moon:2007p11833}), we fitted a spherical black-body component (as in \citealt{Kaplan:2006p11690}) instead of the power law and found a low temperature $T_s=411\pm47$~K and radius $R_s=1288\pm197$~\rsun\ (see Figure~\ref{sed1}, left). 
We note that the \Herschel/PACS data points are well above the model, resulting in a reduced $\chi^{2}$ of 3.95 with 12 degrees of freedom.
We obtained a slightly lower temperature in our fit compared to \citet{Kaplan:2006p11690}.
Following \cite{Moon:2007p11833}, we fixed all the parameters and added a second black-body component to account for this mid-infrared excess. We found a low temperature of 70~K and rather large radius of $\sim$6000~\rsun\ for this second component.
If we thaw the parameters and add instead a component from the stellar wind corresponding to the observed radio flux and with a spectral index of $\alpha=0.6$, the reduced $\chi^{2}$ is improved to 2.16, with a similar temperature and slightly smaller radius (see Figure~\ref{sed1}, right).
Given the low number of degrees of freedom in our fits, the standard deviation from the expected reduced $\chi^{2}$ of 1 is large. Using Gaussian approximation, the width of the $\chi^{2}$ distribution would be 0.4, but it is likely larger in our case (see e.g. \citealt{Andrae:2010p6605}). The reduced $\chi^{2}$ we obtained for models presented in Figure~\ref{sed1} are thus barely acceptable in theory but cannot be completely ruled out. We note that there is a large spread in the residuals for the $JHKs$ bands, and indication of a different curvature than the proposed model (dominated by the stellar photosphere model and the extinction law in those bands).


We then used a classical viscous disk model (as described in Section~\ref{models}) in addition to the CK04 model and obtained acceptable fits (see Figure~\ref{sed2} and Table~\ref{fit_results}). The inner temperature of the disk is consistent with the dust sublimation temperature ($\sim$1\,500~K). 
The addition of a power law corresponding to the stellar wind slightly improved the fit and drastically changed the extension of the disk from $R_{\rm out}\sim 4200$ to $1800$~\rsun. This would correspond to $\sim$25 stellar radii, or $\sim$8~AU. With this model, the residuals for the $JHKs$ bands are not improved. At the outer edge of this classical disk, the temperature drops to $\sim$150~K.


Finally, we fitted the SED with the rimmed disk model defined in Equation~\ref{model_rimmeddisk} (see Section~\ref{models}) and report the results in Figure~\ref{sed3} and Table~\ref{fit_results}. We obtained the best fits with this model and, in particular, flatter residuals for the $JHKs$ bands thanks to the rim component that dominates the contribution to the near-infrared excess. The temperature of the rim is consistent with the dust sublimation temperature as expected for this kind of model (see Section~\ref{models} and  \citealt{Monnier:2005p12460}). The addition of a power law component (stellar wind) again drastically changed the extension of the disk, to reach a similar value to that found for the classical disk, around 1800~\rsun. At the outer edge of this rimmed disk, the temperature drops to $\sim$250~K.

%


\section{Discussion}

Using all the available visible-infrared photometry on \mbox{GX~301--2}, with the addition of \Herschel\ and radio data points for the first time, we fitted spherical and disk-like dust/gas distribution models. We globally found the same reddening, \mbox{\ebv\ $\sim$ 2.0}, and a similar distance, $D\sim3.0$~kpc, with all the different models. This is consistent with previously reported values \citep{Coleiro:2013p10853,Kaper:2006p11663}.
As the stellar radius was fixed to the most likely value found by \cite{Kaper:2006p11663}, the distance is not totally constrained and may be slightly different than the values presented in Table~\ref{fit_results}.


The addition of a power law, used to model the free-free emission from the stellar wind around the HG and detected in radio by \cite{Pestalozzi:2009p12072}, clearly improved the fits for all models. It thus seems likely that this component is required and has some contribution to the mid-infrared flux from the source. 
This contribution could even dominate the 70~\mum\ flux (see Figures~\ref{sed2} and~\ref{sed3}, right).
{Such a free-free emission component has been proposed by \citet{Kaper:2006p11663}, but ruled out by \citet{Kaplan:2006p11690} and thus required confirmation.}

\subsection{Structure of the CS environment}

The model components related to the infrared excess are indicative of the structure of the CS environment of GX~301--2.

We obtain a relatively good fit over five orders of magnitude in wavelength with the simplest model composed of the CK04 stellar photosphere model and a power-law with index $\sim$0.75 (with an appropriate extinction law). 
Such a high index is not ruled out by \cite{Pestalozzi:2009p12072} in the radio range, as their error bars are large, but this index is not consistent with a spherical wind model (index of 0.6 expected).
The global indices from radio to infrared found for similar stars such as P~Cyg seem to reach such a high value.
This may indicate a more complex wind structure, however, a break between radio and near-infrared is clearly possible, with a radio power law index closer to 0.6 and a higher index in the infrared for those objects \citep{Wright:1975p12076}. 
A non-spherical geometry would produce a more flattened radio spectrum. Higher indices could be the consequence of a  lower ionized gas fraction, involving the presence of neutral condensations and possibly the presence of CS dust \citep{Wright:1975p12076}.

It thus seems that the infrared excess in the SED might be explained mostly by the stellar wind.
However, there is a number of features that point towards more complex models.
First, this simple model would contradict the idea that this highly obscured source hosts dust in its CS environment, as revealed by intrinsic silicate absorption lines and continuum emission excess in the infrared (see e.g. \citealt{Kaplan:2006p11690,Moon:2007p11833}).
The \Herschel/PACS data point at 70~\mum\ appears to be more than 3$\sigma$ away from the model, suggesting a curvature in the spectrum. 
Also, the curvature of the data in $JHKs$ is not well followed by the model (see residuals in Figure~\ref{sed0}). 
We thus cannot exclude a break in the mid-infrared with a change of index of the power law slope, or discontinuities as discussed in \cite{Wright:1975p12076}. More data points would be required in order to better constrain the shape of the SED.

We thus give credit to more explicit models that are used to test the possible geometry of the CS gas/dust. 
The spherical distribution models lead to lower fit quality, though they cannot be completely ruled out.
The fits of disk models returned relevant parameter values, {though we note that degeneracies in the fits and a complex extended emission around the source might bring additional uncertainties on the absolute values reported here}.
The size of the dusty disks we derived is about 1\,800~\rsun, (25 stellar radii, or 8~AU) which is comparable to the sizes derived by \cite{Lachaume:2007p11825} for the B[e] star Hen~3--1191 or derived through interferometry for similar objects \citep{Millour:2011p13762}. Compared to the dusty shells observed among most of LBVs \citep[up to a few pc,][]{Clark:2003p12973}, the size is however much smaller. 
The low temperatures we obtained in the disk, which are close to 200~K at the outer edges of the disk, are well below the dust sublimation temperature, indicating that dust can indeed be formed in such a CS environment of GX~301--2.

The rimmed disk model is particularly instructive on the possible structure of the CS environment. It contains an irradiated rim at the inner edge of the disk that efficiently protects the disk from the ionizing radiation of the star.
Following \cite{Lachaume:2007p11825}, the rim radius $R_{\rm rim}$ for a temperature close to the dust sublimation temperature is expected to be around 1\,400~\rsun\ for GX~301--2 in the large gap scenario (their Equation~2, valid for a rimmed disk with optically thin inner gas), which corresponds to the value obtained in our final fit (Figure~\ref{sed3}, right). 
There is a clear degeneracy between $R_{\rm rim}$ and $H_{\rm rim}$ (the latter is fixed in the model). Therefore we cannot give clear constraints to the value of $R_{\rm rim}$.
In the small gap scenario (more similar to a classical disk with optically thick inner gas, see \citealt{Lachaume:2007p11825}), the radius at which the temperature equals the dust sublimation temperature is expected to be $\sim$400~\rsun. However, such temperatures are found at the inner edge of the classical disk in our modelling, close to the star surface (thus significantly below 400~\rsun), arguing against the classical disk model.

For sgB[e] stars, it is generally agreed that a rimmed disk is present around the star, but 
strong [{O}{ I}] emission lines detected in their NIR spectra rather points towards a temperature between 5\,000 and 10\,000~K for the rim \citep{Kraus:2007p12482}. 
In GX~301--2, there are no such lines, and thus, a rim at such a high temperature cannot be present.
Therefore, the rimmed disk solution we obtained, with a rim temperature compatible with the dust sublimation temperature, appears to be plausible.

The size of the dusty disk structure is larger than the neutron star distance to the HG star in GX~301--2, estimated to be 100--200~\rsun, so about twice the stellar radius \cite[e.g.][]{Kaper:2006p11663}. The structure we observe here would thus surround the binary system.
The possible presence of a gas stream close to the periastron of the orbit of the neutron star has been proposed to explain the evolution of the accretion rate, as seen in X-rays \citep{Leahy:2002p13153,Evangelista:2010p13139}. Such a feature would thus not be directly connected, at least spatially, to the CS dusty disk.

\subsection{Comparison with similar objects}

Globally, the CS environment of GX~301--2 appears to be similar to the environment of the obscured HMXB IGR~J16318--4848 hosting a sgB[e] companion star \citep{Chaty:2012p9588} and to sgB[e] stars in general. We thus strengthen the connections between the two systems, that may have undergone a similar evolution that led to the formation of a dense CS environment with a disk hosting dust.
GX~301--2 has known orbital parameters, in particular a significant eccentricity, while there is no evidence for variability for IGR~J16318--4848 and most probably a circular orbit. This difference might be due to a slightly different stage in the evolution after the supernova that formed the neutron star, or to a different initial mass or conditions in the system. 

For another similar source, CI Cam, the geometry of the dusty envelope has been unambiguously determined with long baseline optical interferometry as a torus or a disc (of a few AU in radius) and the hypothesis of a spherical dust shell is totally ruled out \citep{Thureau:2009p13922}. The binary companion would also lie interior to the dusty disc. This structure is similar to the proposed rimmed disk model we tested for GX~301--2. 


{GX~301--2 has been reported to share common properties with 4U 1907+097, which is also a slow pulsar and a O8/O9 supergiant companion with a dense stellar wind \citep{Cox:2005p14024}. It seems to follow a similar evolution track.
This system is a highly obscured HMXB and could be a missing link between supergiant fast X-ray transients (SFXTs) and ordinary accreting pulsars \citep{Doroshenko:2012p14023}. It would be interesting to test the presence of surrounded dust around the system, however, its larger distance and high extinction make it more difficult to study.}

More globally, IGR~J16318--4848 is the prototype of a larger class of obscured X-ray binaries, and as such, this whole class of objects could represent a similar step in the evolution of massive stars in binary systems.
The number of known supergiant HMXBs has dramatically increased over the last decade with new detections using \INTEGRAL. For the HMXBs confirmed by spectral type we now have 49\% of Be HMXBs, 42\% of supergiant HMXBs and 9\% of peculiar HMXBs \citep{Coleiro:2013p13747}, while we had only 4\% of supergiant HMXBs before \INTEGRAL.
Obscured systems hosting a supergiant may be explained by ejections of matter during phases of interaction with the compact object (see e.g. \citealt{Kraus:2010p13136,Wheelwright:2013p12716}). 
This scenario might be related to the formation of a dusty disk around those systems.
More observations and modeling of their infrared excess, and possibly mapping of the CS environment through interferometry, would reveal and confirm the similarities of those objects and allow us to perform population studies.


\subsection{Origin of the dusty disk}

Assuming that a disk of material effectively surrounds GX~301--2, it might have initially formed through the bi-stability mechanism, a scenario proposed for sgB[e] stars \citep{Lamers:1991p13527,Pauldrach:1990p13521}.
This is strengthened by the fact that the surface effective temperature found for the HG star in GX~301--2 is close to 19\,300~K, and was probably higher before its evolution to the HG stage.
With this mechanism it is possible to get a density contrast of about 10 between the equator and the pole. 
Another scenario proposed for the formation of a disk in sgB[e] system requires a high rotational velocity, leading to the formation of an equatorial, wind compressed disk \citep{Bjorkman:1993p13530,Bjorkman:1999p13509}. This scenario might be complementary to the bi-stability mechanism, as shown by \citet{Pelupessy:2000p13780} in order to explain the formation of disks around sgB[e] with observed density contrast of a factor $\sim$100.

In addition to those processes, the presence of a close-by companion (the neutron star in GX~301--2), even when it is much less massive than the primary, could drastically affect a stellar wind. 
It was argued that the formation of a dusty disk could be directly due to the presence of a companion around the massive star and non-conservative mass tranfer \citep{Plets:1995p12556,Clark:2013p13837}.  
Tidal interaction in the binary system could create high density blobs, that would form a rim. This rim could then efficiently protect the disk from the stellar radiation, and enable the formation of dust.
The fact that we favour a rimmed disk model for GX~301--2 appears to give credit to this scenario.
In the same way, some sgB[e] exhibit a disk and the most probable hypothesis is that the accumulation of matter in the equatorial plane is due to the presence of a low mass companion (e.g. \citealt{Millour:2011p13762,Millour:2013p13052}).

\subsection{Evolutionary track}

The evolution of the particular system GX~301--2 has been modeled by \cite{Wellstein:1999p12950} who favoured a scenario with an initial binary system of 25 and 26~\msun, that underwent conservative mass transfer. 
The associated transfer of angular momentum has led to the formation of one of the slowest known pulsar with a wide orbit. It also potentially spun up the HG star, possibly enabling, or strengthening, the mechanism of disk formation around the HG star.
The possible future evolution of the system has been modelled by \citet{Belczynski:2012p13243}, as GX~301--2 is a potential progenitor for a black hole + neutron star binary, and thus among the most promising gravitational wave sources.
However, the expansion of the massive companion will be so rapid that 
the system will probably create a Wolf-Rayet (WR), and then end up with the neutron star sinking into the helium core of the companion star, i.e. a merger.


The disk-like structure of the dust around the HG in \mbox{GX~301--2} is strikingly similar to the structure observed around supergiant stars showing the B[e] phenomenon \citep{Lamers:1998p12450}, suggesting a link between those two classes of objects.
We also note that GX~301--2 (Wray~977) has similar properties to the group of LBV stars \citep{Clark:2012p12954,Clark:2013p13837}, also suggesting a connection between those classes.
It thus appears that there are probably evolutionary links between the HG stage in GX~301--2, sgB[e] systems, LBVs, and obscured HMXBs. 

Forbidden lines have been detected in GX~301--2 \citep{Moon:2007p11833}, though there are differences with the lines more commonly associated to the B[e] phenomenon. 
The inferred temperature for the radiation exciting the [{Ne}{ II}]  and [{Ne}{ III}] lines is hotter than the stellar photospheres, so there can be some contribution from the illumination of hard X-rays from the central compact X-ray source \citep{Moon:2007p11833}.
As [{Ni}{ II}] and [{Fe}{ II}] were detected only in IGR J16318--4848 (that contains a sgB[e]) while [{Fe}{ III}] was only in GX~301--2, this suggests the presence of a slightly harder radiation field in GX~301--2.
One possibility is that the HG stage is similar to the sgB[e] stage for systems with a harder radiative field (e.g. hosting an X-ray source), or that are more massive initially.
It is however more likely that the HG is an evolved sgB[e] (\mbox{sgB[e] $\rightarrow$ HG}) as it now presents a similar CS environment, but a lower stellar effective temperature and larger radius.
As such, a rimmed disk as the one we modeled could be the remnant of an initial sgB[e] disk, with a lower rim temperature, following the lower stellar temperature.

As LBVs generally show large dusty shells that are not observed around GX~301--2, the evolutionary sequence can only be \mbox{HG $\rightarrow$ LBV}, if there is indeed an evolutionary link.
 \citet{Clark:2012p12954} noticed that blue HG yielded physical properties intermediate between blue supergiants and LBVs, such as mass loss rate, wind velocity, stellar luminosity and temperature.
 All this suggests that blue HGs are the immediate descendants of blue supergiants and progenitors of LBVs, for initial masses in the range $\sim$30--60~\msun.


GX~301--2 might represent a peculiar and short-lived stage in the evolution of massive stars in binary systems, that could be represented by the following steps:
\begin{center}
sgB[e] $\rightarrow$ HG $\rightarrow$ LBV $\rightarrow$ WR  $\rightarrow$ SN
\end{center}
This track might be specific to binary systems that become X-ray binaries with an accreting compact object during their evolution, and can be seen as highly obscured X-ray binaries at one point.








\section{Conclusions}

We modeled the optical to radio spectral energy distribution of GX~301--2 and found relatively good fits with different models, all containing a stellar photosphere model and a power law representing the stellar wind. These two components could be sufficient to explain the emission, though they do not explain the presence of dust as reported in previous work. Models including a spherical or disk-like distribution of material can also reproduce the spectral energy distribution. Through comparisons with similar objects, the spherical and classical disk distribution seem less likely to be present in GX~301--2.
The rimmed disk model shows acceptable parameters, and relates to viable mechanisms that would explain the evolution of the system as well as strengthen the connections to other categories of massive stars such as LBVs and sgB[e].




\acknowledgments

We thank the referee for his input and interesting comments.
MS acknowledges founding from the Centre National d’Etudes Spatiales (CNES).
This work is based on observations obtained with MINE -- the Multi-wavelength INTEGRAL NEtwork --, supported by the CNES.
This research made use of Astropy, a community-developed core Python package for Astronomy \citep{AstropyCollaboration:2013p13570}.
This research has made use of the SIMBAD database, operated at CDS, Strasbourg, France.
PACS has been developed by a consortium of institutes led by MPE (Germany) and including UVIE (Austria); KU Leuven, CSL, IMEC (Belgium); CEA, LAM (France); MPIA (Germany); INAF- IFSI/OAA/OAP/OAT, LENS, SISSA (Italy); IAC (Spain). This development has been supported by the funding agencies BMVIT (Austria), ESA-PRODEX (Belgium), CEA/CNES (France), DLR (Germany), ASI/INAF (Italy), and CICYT/MCYT (Spain).



{\it Facilities:} \facility{Herschel}.

\bibliographystyle{apj} 
\bibliography{../../ref.bib}

\end{document}